\documentclass[12pt]{article}
\usepackage{epsf}
\title{\bf Vibrating the $QCD$ string}
\author{
Yu.S.Kalashnikova\thanks{e-mail: yulia@vxitep.itep.ru}, D.S.Kuzmenko
\thanks{e-mail: kuzmenko@vxitep.itep.ru}}
\date{\it Institute of Theoretical and Experimental Physics,\\
117218, Moscow, Russia}
\newcommand{\be} {\begin{equation}}
\newcommand{\ee} {\end{equation}}
\newcommand{\bdm} {\begin{displaymath}}
\newcommand{\edm} {\end{displaymath}}
\newcommand{\bc} {\begin{center}}
\newcommand{\ec} {\end{center}}
\newcommand{\beqa} {\begin{eqnarray}}
\newcommand{\eeqa} {\end{eqnarray}}

\newcommand{\veR}{\mbox{\boldmath${\rm R}$}}
\newcommand{\ver}{\mbox{\boldmath${\rm r}$}}
\newcommand{\vew}{\mbox{\boldmath${\rm w}$}}
\newcommand{\vep}{\mbox{\boldmath${\rm p}$}}
\newcommand{\veL}{\mbox{\boldmath${\rm L}$}}
\newcommand{\vez}{\mbox{\boldmath${\rm z}$}}
\newcommand{\verho}{\mbox{\boldmath${\rm \rho}$}}

\begin{document}

\maketitle

\begin{abstract}
The large-distance behaviour of the adiabatic
hybrid potentials is studied in the framework of the QCD string 
model. The calculated spectra are shown to be the result of
interplay between potential-type longitudinal and string-type
transverse vibrations.
\end{abstract}

General arguments from QCD and lattice data tell that the theory,
even quenched in quarks, possesses nontrivial spectrum, so that
effective degrees of freedom for constituent  glue should be
introduced to describe QCD in the nonperturbative region. As far as
we know the possibility for mesons with gluonic lump to exist was 
first considered in \cite{Okun} in 1976. Modern wisdom tells that the area
law asymptotics for the Wilson loop implies a kind of string to be
developed between quark and antiquark at large distances, and it is
natural to identify the $q\bar q$ system connected by the string in
its ground state with conventional $q\bar q$ meson, while the string
vibrations are responsible for gluonic (hybrid) excitations. This
picture, though physically appealing, does not follow directly from
the QCD, and one relies upon models to describe these excitations.
There are two main ideas on how to construct such models. One is to
consider point-like gluons confined by some potential-type force
\cite{H&M,S&S}, and another is to introduce string phonons 
\cite{fluxtube}.

In principle the best way to discriminate between these two
possibilities is to compare predictions with experimental data on
hybrid mesons. Indeed, there  is a lot of indications that hybrid
mesons are already found, but the conclusive evidences have
never been presented, nor have alternative explanations been
completely excluded \cite{data}.

On the other hand, lattice calculations are now accurate enough to
provide reliable data on the properties of soft glue and to check the
model predictions. In this regard recent measurements \cite{lattice} of
adiabatic hybrid potentials are of particular interest. These
simulations measure the spectrum of  glue in the presence of static
quark and antiquark separated by some distance $R$. Not only these
potentials enter heavy hybrid mass estimations in the
Born-Oppenheimer approximation. The large $R$ limit is important
\begin{it} per se, \end{it}
as the formation of confining string is expected at large
distances, and direct measurements of string fluctuations become
available. It is our purpose to investigate the large-distance
behaviour of adiabatic potentials in order to establish what kind of
the effective string degrees of freedom are excited at large distances.   

We perform these studies in the framework of the QCD string model. 
This model deals with quarks and point-like gluons propagating in the
confining QCD vacuum, and is based on Vacuum Background Correlators  
method \cite{Lisbon}. 
The QCD string 
model was successfully applied to conventional mesons \cite{DKS}, 
hybrids \cite{Simhybrids,hybrids,Lisbon}, glueballs \cite{KS} and gluelump
(gluon bound to the static adjoint source) \cite{S}. 

The QCD string model for gluons is derived from the perturbation theory
in the nonperturbative background, developed in \cite{Pert}. This
formalism allows to introduce constituent (valence) gluons as
perturbations at the confining background. The latter is given by the
set of gauge-invariant field strength correlators responsible for the
area law. The main feature of this approach is that, in contrast to the
above-mentioned models, here one is able to distingiush clearly
between confining gluonic field configurations and confined valence
gluons.

The starting point is the Green function for the gluon propagating in
the given background field $B_\mu$ \cite{Pert}:
\be
G_{\mu\nu} (x,y) = (D^2(B)\delta_{\mu\nu}+ 2ig F_{\mu\nu} (B))^{-1},
\ee
where covariant derivative $D_{\lambda}^{ca}(B)$ is
\be
D_{\lambda}^{ca}(B)=\delta_{ca} \partial_{\lambda}+g f^{cba} B^b_{\lambda}.
\ee

The term proportional to $F_{\mu\nu}(B)$ is responsible for the gluon
spin interaction; in these first studies we neglect it, as it can be
treated as perturbation \cite{KS,S}. The next step is to use
Feynman-Schwinger representation for the quark-antiquark-gluon Green
function \cite{hybrids}, which is reduced in the case of  static quark
and antiquark to the form
\be
G(x_g, y_g)= \int^\infty_0 ds \int Dz_g \exp (-K_g) \langle
{\cal W}\rangle_B,
\ee
where $K_g=\frac14 \int^s_0\dot z^2_g(\tau) d\tau$, and all the
dependence on the vacuum gluonic field $B_\mu$ is contained in the
Wilson loop
\be
{\cal W}=\mathrm{Sp}(\lambda_a\Phi_q\lambda_b \Phi_{\bar
q})\Phi^{ab}_{\Gamma_g} (y_g, x_g).
\ee
Here $\Phi_q$ and $\Phi_{\bar q} $ are   parallel transporters
\be
\Phi_q= P\exp ig \int^{x_q}_{y_q} B_\mu(z_q) dz_{q\mu},~~
\Phi_{\bar q}= P\exp ig \int^{y_{\bar q}}_{x_{\bar q}}
B_\mu(z_{\bar q}) dz_{\bar q\mu}
\ee
with integration in (5) along the classical trajectories
$z_{q\mu}=(\tau, \frac{\veR}{2})$ and $z_{\bar q \mu}= (\tau,
-\frac{\veR}{2})$ of static quark and antiquark, $P$ means path ordering,
and
\be
\Phi^{ab}_{\Gamma_g} (y_g, x_g) = ( P\exp ig \int_{\Gamma_g} B_\mu
(z_g) dz_{g\mu})^{ab},
\ee
$a,b$ are adjoint colour indices, $\lambda_a$ are Gell-Mann matrices
and the contour $\Gamma_g$ runs over the gluon trajectory $z_g$.

The main assumption of the QCD string model is the minimal area law
for the Wilson loop average, which yields for the configuration
(4) the form \cite{hybrids}
\be
\langle {\cal W}\rangle_B = \frac{N_c^2-1}{2}\exp (-\sigma(S_1+S_2)),
\ee
where  $S_1$ and $S_2$ are the minimal areas inside the contours
formed by quark and gluon and antiquark and gluon trajectories
correspondingly, and $\sigma $ is  the string tension.

With the form (7) for $\langle {\cal W}\rangle_B$ the action of the
system can be immediately read out of the representation (3):
$$
A=\int^T_0 d\tau \left \{ -\frac{\mu}{2}+\frac{\mu\dot r^2}{2}- \sigma
\int^1_0 d\beta_1
\sqrt{(\dot w_1 w_1')^2-\dot w_1^2w^{'2}_1}-\right.
$$
\be
\left.-\sigma\int^1_0 d\beta_2
\sqrt{(\dot w_2 w_2')^2-\dot w_2^2w^{'2}_2}\right \},
\ee
where the minimal surface $S_1$ and $S_2$ are parametrized by the
coordinates $w_{i\mu}(\tau,\beta_i) ,~ i=1,2,~
\dot w_{i\mu}= \frac{\partial w_{i\mu}}{\partial \tau},~
w'_{i\mu}= \frac{\partial w_{i\mu}}{\partial \beta_i}.$

In what follows the straight-line ansatz is chosen for the minimal
surface:
\be
w_{io}=\tau,~~~\vew_{1,2} = \pm (1-\beta) \frac{\veR}{2}+\beta
\ver.
\ee
The quantity $\mu=\mu(\tau)$ in the  expression  (8) for the
action is the so-called einbein field \cite{einbein}; here
one is forced to introduce it, as  it is the only way to obtain
meaningful dynamics for the massless particle. Moreover, we
introduce another set of einbein fields,
$\nu_i=\nu_i(\tau,\beta_i)$ to get rid of Nambu-Goto square roots
in (8) \cite{DKS}.
The resulting Lagrangian takes the form
$$
L=-\frac{\mu}{2}+\frac{\mu\dot r^2}{2}-
\int^1_0 d\beta_1\frac{\sigma^2 r^2_1}{2\nu_1}-\int^1_0
d\beta_1\frac{\nu_1}{2}(1-\beta_1^2 l_1^2)-
$$
$$
-\int^1_0 d\beta_2\frac{\sigma^2 r^2_2}{2\nu_2}-\int^1_0
d\beta_2\frac{\nu_2}{2}(1-\beta_2^2 l_2^2),
$$
\be
l^2_{1,2}=\dot r^2-\frac{1}{r^2_{1,2}}(\ver_{1,2}\dot{\ver})^2,~~
\ver_{1,2}=\ver\pm \frac{\veR}{2}.
\ee

It is clear from Eq.(10) that the einbein field $\mu$ can be
treated as the kinetic energy of the constituent gluon, and the einbeins
$\nu_i(\tau,\beta_i)$ describe the energy density distribution along the
string. These quantities are not introduced by hand, but are calculated
in the presented formalism.
Indeed, as no time derivatives of the  einbeins enter the
Lagrangian  (10), it describes the constrained system, with the
equations of motion
\be
\frac{\partial L}{\partial\mu}=0,
~~\frac{\delta L}{\delta \nu_i(\beta_i)}=0
\ee
playing the role of second-class constraints.

Now one obtains the Hamiltonian $H=\vep\dot{\ver}-L$ with the
result
\be
H=H_0+\frac{\mu}{2}
+\int^1_0 d\beta_1\frac{\sigma^2 r^2_1}{2\nu_1}
+\int^1_0
d\beta_2\frac{\sigma^2 r^2_2}{2\nu_2}+\int^1_0
d\beta_1\frac{\nu_1}{2}+
\int^1_0
d\beta_2\frac{\nu_2}{2},
\ee

$$
H_0=\frac{p^2}{2(\mu+J_1+J_2)}+
$$
$$
\frac{1}{2\Delta (\mu+J_1+J_2)}
\left\{ \frac{(\vep\ver_1)^2}{r^2_1}J_1(\mu+J_1)+
\frac{(\vep\ver_2)^2}{r^2_2}J_2(\mu+J_2)+\right.
$$
\be
\left.\frac{2J_1J_2}{r^2_1r^2_2}(\ver_1\ver_2)(\vep\ver_1)(\vep\ver_2)
\right\}
\ee

$$
\Delta=(\mu +J_1) (\mu+J_2) -
J_1J_2\frac{(\ver_1\ver_2)^2}{r^2_1r^2_2},~~
J_i=\int^1_0 d\beta_i\beta_i^2\nu_i(\beta_i),~~ i=1,2.
$$

As we deal with the constrained system, the extra variables $\mu$
and $\nu_{1,2}$ should be excluded by means of the conditions
\be
\frac{\partial H}{\partial\mu}=0,~~
\frac{\delta H}{\delta \nu_i(\beta_i)}=0
\ee
before quantization; the extrema of the einbeins should be found
from the equations (14) and substituted into the Hamiltonian. Such
procedure is hardly possible analytically with the complicated
structure (12), (13) even at the  classical level, and
after quantization these extremal values of einbeins would become
nonlinear operator functions of coordinates and momenta with
inevitable ordering problems arising. In what follows we use the
approximation which treats $\mu$ and $\nu_i$ as $c$-number
variational parameters. We find the eigenvalues of the Hamiltonian
(12) as functions of $\mu$ and $\nu_i$ and minimize them with
respect to einbeins to obtain the physical spectrum. Such einbein
method
works surprisingly well in the QCD string model calculations, with the
accuracy of about 5-10\% for the ground state \cite{rotstring}.

Even with this simplifying assumption the problem
remains complicated due to the presence  of the terms $J_{1,2}$
responsible for the string inertia. Suppose for a moment that one can
neglect these terms in the kinetic energy (13). Then the
Hamiltonian takes the form \cite{hybrids,Lisbon}
\be
H=\frac{p^2}{2\mu}+\frac{\mu}{2}+
\int^1_0d\beta_1\frac{\sigma^2r^2_1}{2\nu_1}+
\int^1_0d\beta_2\frac{\sigma^2r^2_2}{2\nu_2}+
\int^1_0d\beta_1\frac{\nu_1}{2}+
\int^1_0d\beta_2\frac{\nu_2}{2},
\ee
which allows to eliminate einbeins and to arrive at the potential
model Hamiltonian
\be
H=\sqrt{p^2}+ \sigma r_1+\sigma r_2.
\ee

Let us now estimate whether the neglect of string inertia is
justified. To this end we find the spectrum of the Hamiltonian (15),
(16) using the einbein method described above. It is given by
the set of equations
$$
E_n(R) =\mu_n(R) +\frac{4(n+\frac32)^2\sigma^2}{\mu^3_n(R)}
$$
\be
16\sigma^2(n+\frac32)^4= \mu^4_n(R)(4(n+\frac32)^2+R^2\mu^2_n(R))
\ee
with $\nu_i$ independent of $\beta_i$:
\be
\nu_{1n}(R)= \nu_{2n}(R) =\frac{2(n+\frac32)^2\sigma^2}{\mu^3_n (R)},
\ee
where $n=n_z+n_\rho+\Lambda,~~ \Lambda=\left
\vert\frac{\veL\veR}{R}\right\vert$ is the projection of orbital
momentum onto $z$ axis, $\vez\parallel\veR$. Note that while the
angular momentum is not  conserved in the exact Hamiltonian
(16), it is a good quantum number in the approximate einbein
method: we have compared the spectrum of exact and einbein-field
Hamiltonian and have found that angular momentum is conserved in the
potential problem (16) within better than 5\% accuracy. The same
phenomenon is observed in the constituent gluon model \cite{S&S}, 
and is the consequence of linear potential confinement.
 
Consider first the small $R, R\ll 1/\sqrt{\sigma}$, limit of the
system (17):
\be
E_n(R) =2^{3/2}\sigma^{1/2}
(n+\frac32)^{1/2}+\frac{\sigma^{3/2}R^2}{2^{3/2}(n+\frac32)^{1/2}},
\ee
$$
\mu_n(R) =2^{1/2}\sigma^{1/2}
(n+\frac32)^{1/2}-\frac{\sigma^{3/2}R^2}
{(n+\frac32)^{1/2}
2^{5/2}},
$$

$$
\nu_{1,2 n}(R) =
\frac{(n+\frac32)^{1/2}\sigma^{1/2}}
{2^{1/2}}+\frac{3
\sigma^{3/2}R^2}{2^{7/2}
(n+\frac32)^{1/2}}.
$$

The last line in (19) yields $J_{1,2}/\mu\approx \frac16$. The
situation here is similar to the one  in the light quark, glueball
and gluelump QCD string calculations: the correction due to string
inertia is sizeable but not large, and can be taken into account as
perturbation \cite{KS,S}. Note that it is the regime of small $R$ which
is relevant to the heavy hybrid mass estimations \cite{heavyhybr}: 
the average distance between heavy quark and antiquark is small,
$\langle R^2\rangle\ll 1/\sigma$, so that $Q\bar Q$ pair
resides in the oscillator adiabatic potential which, in the einbein
method, is given by Eq. (19).

The situation changes drastically for the case of large $R,~ R\gg
1/\sqrt{\sigma}$. Now gluon enjoys small oscillation motion, and one
has
\be
E_n(R)=\sigma
R+\frac{3}{2^{1/3}}\sigma^{1/3}\frac{(n+\frac32)^{2/3}}{R^{1/3}},
\ee
$$
\mu_n(R)=\frac{4\sigma^{1/3}(n+\frac32)^{2/3}}{R^{1/3}},~~
\nu_{1,2 n}(R)=\frac{\sigma R}{2},
$$
displaying $(\frac{\sigma}{R})^{1/3}$ subleading behaviour typical 
for linear potential confinement at large distances \cite{Lisbon}. 
Nevertheless, in this case $J_{1,2}=\frac16 \sigma R \gg\mu_n$, so that
the potential regime is unadequate at large $R$.

To get more insight into what happens at the intermediate and 
large distances  we consider the quasiclassical limit of 
large $\Lambda$, where only rotations around $z$ axis are taken 
into account:
$$
H=\frac{\Lambda^2}{2\rho^2(J_1+J_2)}+
$$
$$
\frac{\sigma^2}{2}(\rho^2+(z+\frac{R}{2})^2)\int\frac{d\beta_1}{\nu_1}
+\frac{\sigma^2}{2}(\rho^2+ (z-\frac{R}{2})^2)
\int\frac{d\beta_2}{\nu_2}+
$$
\be
\int^1_0 d\beta_1\frac{\nu_1}{2}
+\int^1_0 d\beta_2\frac{\nu_2}{2}.
\ee
As no momenta $p_z$ and $\frac{\vep\verho}{\rho}$ enter the
Hamiltonian, the system stabilizes itself at the points $z_0$ and
$\rho_0$ given by the conditions
\be
\frac{\partial H}{\partial z}=0,~~
\frac{\partial H}{\partial \rho}=0.
\ee
Combining Eq.(22) with the second condition of Eq.(14)
one arrives at the following expressions:
\be
z_0=0,~~ \rho_0=\frac{\Lambda}{2\sigma\sqrt{Ja}},~~ \nu_1(\beta)=
\nu_2(\beta) =\nu(\beta),
\ee
where
\be
J=\int^1_0d\beta \beta^2\nu(\beta),~~ a=\int^1_0\frac{d\beta}{\nu(\beta)}
\ee
and the function $\nu(\beta)$ is given by
\be
\nu(\beta)=\frac{\sqrt{A}}{\sqrt{1-B\beta^2}},~~
A=\frac{\sigma^2R^2}{4}+\frac{\Lambda\sigma}{2\sqrt{aJ}},~~
B=\frac{\Lambda\sigma}{2J}\sqrt{\frac{a}{J}}.
\ee
Substituting the form (25) into eqs. (24) one finds the
expression for the energy
\be
E= 2\sigma^{1/2}\Lambda^{1/2}\mathrm{arcsin}\sqrt{B}\frac{ \{
\mathrm{arcsin}\sqrt{B}+\sqrt{B(1-B)}\}^{1/4}}
{\{\mathrm{arcsin}\sqrt{B}-\sqrt{B(1-B)}\}^{3/4}},
\ee
$$
2\Lambda B^{3/2} \sqrt{1-B}= 
$$
$$
\frac{\sigma R^2}{4}\{
\mathrm{arcsin}\sqrt{B}+\sqrt{B(1-B)}\}^{1/2}
\{\mathrm{arcsin}\sqrt{B}-\sqrt{B(1-B)}\}^{3/2},
$$
with the large $R$ limit of (26) given by
\be
E(R)=\sigma R+2\sqrt{3}\frac{\Lambda}{R}.
\ee
Here we have $1/R$ subleading behaviour typical for naive Nambu-Goto 
string models. For example, the flux-tube model \cite{fluxtube}
predicts
\be
E(R) =\sigma R+\frac{\pi\Lambda}{R}
\ee
in the small-oscillation approximation. The energy curve
(26) is shown at Fig. 1 together with the flux-tube
(28) and potential-regime curve (17) for
$n_z=n_\rho=0$ and $\Lambda=1,2,3$. The large $R$ limit of
the quasiclassical regime (26) is very close to the flux tube one and
deviates substantially from the potential regime, while at small $R$
unphysical divergent $1/R$ behaviour is absent.

\begin{figure}[t]
\epsfxsize=11.6cm
\centering
\epsfbox{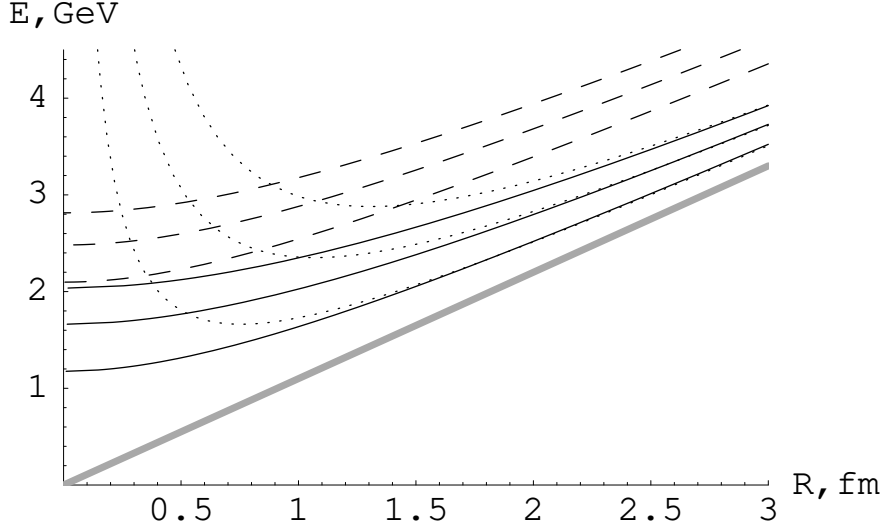}
\caption{Adiabatic hybrid potentials in various regimes. Quasiclassical
 (solid line), potential (dashed), and flux-tube
(dotted) curves for $n_z=n_\rho=0$ and $\Lambda=1,2,3$. The lowest curve is
$\sigma R$. $\sigma = 0.22$GeV$^2$.}
\end{figure}

The case of large $R$ can be treated directly in the full
Hamiltonian (12), which in the small-oscillation
limit takes the form
$$
H=\frac{\mu}{2}+\frac{p^2_z}{2\mu}
+\frac{p_{\perp}^2}{2(\mu+J_1+J_2)}
$$
$$
+\sigma^2(\rho^2+(z+\frac{R}{2})^2)
\int^1_0\frac{d\beta_1}{2\nu_1}+
\sigma^2(\rho^2+(z-\frac{R}{2})^2)
\int^1_0\frac{d\beta_2}{2\nu_2}
$$
\be
+\int^1_0{d\beta_1}\frac{\nu_1}{2}
+
\int^1_0{d\beta_2}\frac{\nu_2}{2},
\ee
displaying two different kinds of string excitations, along the $z$
axis and in the transverse direction. Indeed, for large $R$ one
neglects the contribution of $\mu$ in the third term of (29)
because the extremal values of $\nu_{1,2}$
are $\frac{\sigma R}{2}$. Then the oscillations in the longitudinal
and transverse directions become uncoupled, and one has
\be
E_n(R)=\sigma
R+\frac{3}{2^{1/3}}\frac{\sigma^{1/3}(n_z+\frac{1}{2})^{2/3}}{R^{1/3}}+
\frac{2\cdot 3^{1/2}}{R}(n_\rho+\Lambda+1).
\ee
The regime $\sim (\frac{\sigma}{R})^{1/3}$ is established at large $R$,
but at the intermediate distances there are sizeble corrections from the
string regime $\sim \Lambda/R$, as it is seen from Fig. 2.  

\begin{figure}[t]
\epsfxsize=16.5cm
\centering
\epsfbox{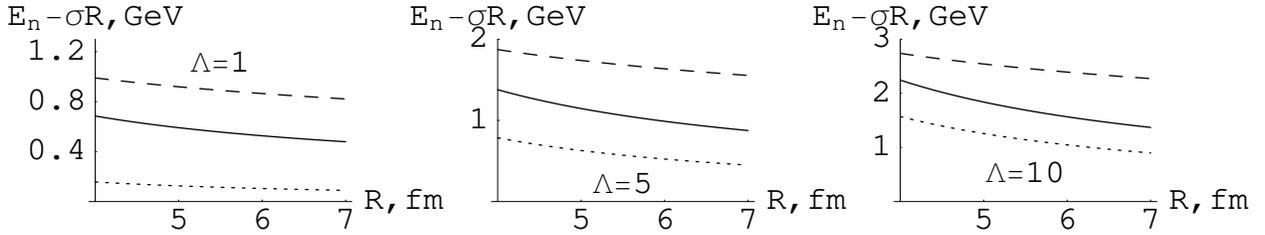}
\caption{Corrections to linear behaviour of potentials. QCD string (solid
line), potential (dashed), and flux-tube (dotted) curves; $n_z=n_{\rho}=0$;
$\sigma = 0.22$GeV$^2$.}
\end{figure}

As we have not considered the spin of the gluon, we are not in the
position yet to compare our predictions with lattice results
\cite{lattice}. Nevertheless, some preliminary conclusions can be drawn.
For separations less than 2 fm the measured energies \cite{lattice}
lie much below Nambu-Goto curves (28).
There is no universal Nambu-Goto behaviour
even for $R$ as large as 4 fm. The QCD string model is able to describe 
both these features: at small separations the potential confinement 
regime dominates, while at large distances the situation is more
complicated. Indeed, there is the contribution of the string-type gaps
(27) which
are due to transverse vibrations of the string, but the dominant
subleading behaviour is defined by potential-type longitudinal motion.
In particular,
even for quasiclassically large values of $\Lambda$ there exists the
contribution of oscillations in the longitudinal direction (second
term in (30)).

Such peculiar behaviour displays the most pronounced difference 
between the given approach and other models of constituent glue.
In contrast to phonon-type models, 
the QCD string vibrations are caused by point-like valence gluon,
but, in contrast to potential models, the confining force follows
from minimal area law, giving rise, at large distances, both to
longitudinal vibrations with potential-type $\sim (\frac{\sigma}{r})
^{1/3}$ 
dominant subleading behaviour and to the transverse vibrations
with string-type $\sim \Lambda/R$ subleading behaviour, which 
could be responsible for the observed $\Lambda$ dependence. The
full QCD string calculations with gluon spin involved will provide,    
if confirmed by the lattice data, the decisive evidence in favour 
of the QCD string model of valence glue.  

We are grateful to Yu.A.Simonov for useful discussions.
The support of INTAS-RFFI 97-0232, RFFI 00-02-17836 and
00-15-96786 grants is acknowledged.

\end{document}